\documentclass[11pt]{article}
\usepackage{amsmath, amssymb}
\usepackage{verbatim}
\usepackage{latexsym}
\usepackage{euscript}
\usepackage{amsfonts}
\usepackage{verbatim}
\def\be{\begin{eqnarray}}
\def\ee{\end{eqnarray}}
\def\0{\nonumber}

\newcommand\EE{\EuScript{E}}
\newcommand\EF{\EuScript{F}}
\newcommand\EK{\EuScript{K}}
\newcommand\EH{\EuScript{H}}
\usepackage[]{array}
\usepackage[]{amsmath}
\usepackage[]{amssymb}
\usepackage[]{mathrsfs}
\usepackage[]{tensor}
\usepackage[symbol]{footmisc}
\newcommand{\del}{\partial}
\newcommand{\tr}{\textrm{tr}}

\newcommand{\e}{\varepsilon}
 
\begin{document}

\begin{flushright}
SISSA/12/2013/FISI\\  
hep-th/1304.2159
\end{flushright}
\vskip 2cm
\begin{center}

{\LARGE {\bf  Generalized states in SFT}}
 
\vskip 1cm

{\large L. Bonora  and S. Giaccari   }

{}~\\

\quad \\

{\em ~ International School for Advanced Studies (SISSA),}\\

{\em  Via Bonomea 265, 34136 Trieste, Italy and INFN, Sezione di Trieste}\\

{\tt bonora@sissa.it, giaccari@sissa.it}

 \end{center}

\vskip 2cm {\bf Abstract.}
{The search for analytic solutions in open string fields theory \`a la Witten often meets with singular expressions, which need an adequate mathematical formalism to be interpreted. In this paper we discuss this problem and propose a way to resolve the related ambiguities. Our claim is that a correct interpretation requires a formalism similar to distribution theory in functional analysis. To this end we concretely construct a locally convex space of test string states together with the dual space of functionals. We show that the above suspicious expressions can be identified with well defined elements of the dual.}
\eject

\section{Introduction}

The search for exact (analytic) solutions in the SFT \`a la Witten, \cite{Witten1}, has been characterized not only by considerable successes (see \cite{Schnabl05,Okawa1,ErlerSchnabl,RZ06,ORZ,Fuchs0} and \cite{ KORZ,Kiermaier:2007vu,Fuchs3,Lee:2007ns,Kwon:2008ap,Erler07,Okawa2,Okawa3,Kiermaier:2007ki,Erler:2007rh,Arefeva09,Arefeva10,Arroyo:2010fq,Zeze:2010jv,Zeze:2010sr,Arroyo:2010sy,Murata,Murata2,Baba,Hata,Masuda1,Masuda2}, as well as \cite{BMT,BGT1,BGT2,BGT3,EM,EM2,EM3} and the reviews \cite{Fuchs4,Schnabl:2010tb}), but also by remarkable improvements in the mathematical language of the theory. An example of the latter is the astonishingly simple form taken by such analytic solutions thanks especially to the so called $K,B,c$ algebra. 
However in time it has become evident that these new tools have to be handled with more care than one may have initially imagined, due to the presence of possible singularities, and are in need of a more accurate mathematical formalization. The purpose of this paper is to start the analysis of these problems and put forward a proposal for their solution. Our main claim is that these obstacles can be overcome with the use of an appropriate formalism, inspired by the theory of distributions in ordinary functional analysis.  To avoid being generic we will discuss a very concrete problem. To start with let us illustrate it. 

In SFT one faces the problem of defining the inverse of objects (string fields) like $K$ or $K+\phi$, where 
\be 
K=\frac{\pi} 2 K_1^L | I\rangle,\quad\quad \phi=\phi\left(\frac12\right)|I\rangle\0
\ee
For $K_1^L$, see Appendix, 
$\phi(z)$ is a matter field evaluated in the arctan frame. The inverse of these two string fields is usually interpreted via the Schwinger representation 
\be
\frac 1{K}= \int_0^\infty dt\, e^{-tK},\quad\quad \frac 1{K+\phi}= \int_0^\infty dt\, e^{-t(K+\phi)}\label{Schwinger}
\ee
These expressions are thought to be singular, on the basis of the presence in $|I\rangle$ of the zero mode of $K_1^L$ and ${\cal K}=K_1^L+\phi\left(\frac 12\right) $, respectively, which renders the representation (\ref{Schwinger}) infinite, in both cases. On the contrary, for instance, it is believed that 
 \be
\frac 1{K+1}= \int_0^\infty dt\, e^{-t(K+1)}\label{invK+1}
\ee
is well defined due to the factor $e^{-t}$ in the Schwinger representation. 

These assumptions are based on the fact that $\lim_{t\to\infty} e^{-t K}$ is a representation of the sliver, which is supposed to be a well defined object in the $K,B,c$ algebra (a similar argument is believed to hold also in the case of $K+\phi$). As a consequence negative eigenvalues of $K_1^L$ or ${\cal K}$, if any, are ineffective. It should be noticed that the above singularities can be `localized', via the Schwinger representation, at $t\to\infty$. Therefore they are singularities in the auxiliary $t$ space, which represents the string world-sheet boundary, not in space-time. So they simply call for a regularization.

The simplest possible regularization for (\ref{Schwinger}) is  
\be
\lim_{\e\to 0}\frac 1{K+\e} \quad\quad {\rm and} \quad\quad \lim_{\e\to 0} \frac 1{K+\phi+\e} \label{regSchwinger}
\ee 
for positive $\e$. We would like to notice however that in all the above limits the topology is unspecified. The purpose of this paper is to try to clarify this question. We will see in particular that (\ref{regSchwinger}) alone is too simple-minded.

One possible way to shed light on the previous assumptions, is to analyse the spectrum of the operators $K_1^L$ and ${\cal K}$, which act on $|I\rangle$. Now, both $K_1^L$ and ${\cal K}$ are symmetric operators whose spectrum lies on the real axis. Knowing in detail their spectrum would allow us to analyse the convergence of the Schwinger representation in (\ref{Schwinger},\ref{invK+1}) and (\ref{regSchwinger}). As we show in the Appendix, however, this way seems to be rather impervious. Thus what remains for us is to rely on correlators.

A related problem is represented by the meaning of limiting expressions such as
\be 
\lim_{\e\to 0} \frac {\e}{K+\e} \label{first}
\ee
and
\be 
\lim_{\e\to 0} \frac {\e}{K+\phi+\e} \label{second}
\ee
These objects, which are used in constructing and discussing analytic solutions of the SFT equation of motion, are not altogether new in the literature. They bring together two old problems. What is new is that these old problems appear
simultaneously in a new context, that of SFT. First we remark that both (\ref{first}) and 
(\ref{second}) have `pointlike support', if any. Both in fact can be nonzero only in correspondence with one particular value taken by $K$ or $K+\phi$, which corresponds to the zero mode of $K_1^L$, ${\cal K}$, respectively. These expressions are somehow the analog of objects like
\be
\lim_{\e\to 0} \frac {\e}{x^2+\e} \label{first'}
\ee
defined on the real line, which has support, if any, at $x=0$. It is well-known that (\ref{first'})  can be interpreted as an ordinary distribution. Distributions may be ordinary functions, but the interesting thing about them is that they allow us to define objects which are almost functions but not quite. In order to be able to evaluate them one has first to define a space of regular (test) functions with its topology and define a rule in order to evaluate the distributions on such test functions. Distributions are in fact linear functionals over the test function space. Therefore they belong to the dual space of the latter. The limit (\ref{first'}) is actually taken in the dual space, which must therefore be equipped with an adequate topology.

Thus formulae like (\ref{Schwinger},\ref{first},\ref{second}) are to be considered empirical at least as long as the right mathematical context in which they have to be understood is not clearly defined. That is, like for ordinary distributions, we have to define the appropriate couple of 
dual spaces. But, of course, (\ref{Schwinger},\ref{first}) and
(\ref{second}) are no ordinary distributions. The role of points in space for ordinary distributions is played here by states in first quantized string theory. This is the second problem alluded to above. A similar one was met in quantum mechanics and solved long ago by means of Gelfand triples (or rigged Hilbert spaces). Let us consider the simple example of a 1d nonrelativistic particle on the real line. The quantum description is obtained  by solving the Schroedinger equation with suitable convergence properties at infinity, so that the wave-function $\psi$ is square integrable on the real line. The completion of the space of such functions leads to a Hilbert space $\boldsymbol {\cal H}$. However the position and momentum operators,
and polynomials thereof, are not well defined on all functions of the Hilbert space; for this to be the case one has to single out the subspace $\boldsymbol \Phi$ of smooth functions. This is not enough because neither in $\boldsymbol \Phi$ nor in $\boldsymbol {\cal H}$ can we accommodate
the eigenfunctions of the position and momentum operator. But, the latter nicely fit
in the dual space $\boldsymbol \Phi'$ of $\boldsymbol \Phi$, i.e. in the space of linear continuous functionals on $\boldsymbol \Phi$. An important aspect is that the topology of  $\boldsymbol \Phi$ has to be chosen in such a way that the application of the position and momentum operators are continuous, which requires a stronger topology than the Hilbert space 
topology of $\boldsymbol {\cal H}$. As a consequence the dual space ${\boldsymbol \Phi'}$ is larger. We have in fact the inclusion
\be
{\boldsymbol \Phi}\subset {\boldsymbol {\cal H}}\subset {\boldsymbol \Phi'}\0
\ee
This is called rigged Hilbert space or Gelfand triple. The elements of $\boldsymbol \Phi'$ are
often called `distributions' too.  

The point advocated in this paper 
is that a similar construction has to be envisaged in order to correctly interpret (\ref{Schwinger},\ref{regSchwinger},\ref{first}) and (\ref{second}). The general structure is always the same. We have a space of `regular' objects,
say ${\boldsymbol{\cal R}}$ with a suitable topology, its completion $\bar{\boldsymbol{\cal R}}$ and the dual space
${\boldsymbol{\cal R}'}$ of `distributions', with the inclusion
\be
{\boldsymbol{\cal R}}\subset \bar {\boldsymbol{\cal R}}\subset {\boldsymbol{\cal R}'}\label{inclusion}
\ee
and with the duality rule in order to evaluate elements of ${\boldsymbol{\cal R}'}$ on elements of  ${\boldsymbol{\cal R}}$. 

In particular, {\it we will say that a `distribution' is zero whenever it vanishes when evaluated on all the elements of  ${\boldsymbol{\cal R}}$}.  

This paper is a first attempt to define a suitable mathematical framework, i.e. to define appropriate spaces ${\boldsymbol{\cal R}}$ and  ${\boldsymbol{\cal R}}'$, that allow us to interpret formulas like the ones above which are met in SFT.

The paper is organized as follows. The next section contains a pedagogical reminder of distribution theory in functional analysis. It is intended to provide a guideline for the subsequent mathematical developments. For this reason it contains examples which will be used as paradigms in the sequel. In section 3 we outline the same approach in a Fock space appropriate for string theory. In section 4 we introduce the main mathematical problem of this paper, presented in the context where it has appeared, that is the search for analytic lump solutions. In section 5 we construct a set of test (string) states and in section 6 we show that it forms a locally convex topological vector space, where it is possible to define a weak and strong topology. The dual (either weak or strong) of this topological vector space will be our sought for space of `generalized states' or `distributions'. In section 7 we discuss the application of this formalism to the problem of section 4: the previously introduced `principally regularized' are seen to be appropriate to solve it.

\section{Ordinary distributions}

{\it Distributions} or {\it generalized functions} is a branch of functional analysis which deals with objects that are {\it almost functions} but not quite: they are {\it irregular}, but in a controllable way. In which sense this must be understood is explained by the following (qualitatively expressed) general result:
\vskip 0.5cm
{\it Distributions can be reduced to finite order derivatives of locally integrable functions}.
\vskip 0.5cm
An example is the Dirac delta function on the real line, which can be viewed as the derivative of the Heaviside step function; the latter is of course locally integrable. Such irregular functions need to be regularized. Distribution theory provides a `canonical' way to regularize them. The reason why we need such objects is that they appear
in many physical problems, which cannot be described in terms of ordinary functions.  

This section is a short review of the main properties of ordinary distributions. It is intended to be a guideline for the later developments. The expert reader can skip it.  

In order to be able to carefully define distributions one has to view them as linear continuous functionals of a topological vector space formed by ordinary continuous differentiable  functions with good convergence properties at infinity, hereafter called {\it test functions}. Defining a functional means defining a (linear) rule that associates a number to any test function. This rule always consists of a Lebesgue integral.

The (vector) space of test functions must be topological, because a topology is necessary in order to tell what
test functions are close to one another and what are not. In this way we can define continuity for functionals. Also the dual space, the space of functionals, must have a topology, because we want to be able to take limits
of distributions. 

\subsection{Example of test function spaces}

Test functions will be denoted by Greek letters $\varphi,\psi,...$. A space of test functions is, for instance, ${\mathbf K}(a)$: it is formed by all the functions on the real line which are infinitely differentiable and have support inside the interval $|x|\leq a$. A linear combination of them has still the same characteristics, therefore ${\mathbf K}(a)$ is a vector space.  ${\mathbf K}(a)$ is also a topological vector space, but its topology is quite nontrivial. In fact it is not a normed vector space, that is a space characterized by a norm $||\cdot||$, but it is a {\it countably normed vector space}, that is it is characterized by an infinite sequence of norms.
They are defined as follows
\be 
||\varphi||_p= {\rm max}_{|x|\leq a} \{ \varphi(x),\varphi'(x),\ldots,\varphi^{(p)}(x)\}, \quad p=0,1,2,\dots\0
\ee
We have of course $||\varphi||_p\leq ||\varphi||_{p+1}$.
Since  ${\mathbf K}(a)$  is a vector space its topology is defined by a set of neighborhoods of 0. The latter are given by
\be
U_{p,\epsilon}= \{||\varphi||_p<\epsilon\}\0
\ee
One can prove that this defines a topology. 

We can complete ${\mathbf K}(a)$  with respect to one of these norms, say $||\cdot||_p$. In such a way we get 
$\overline{{\mathbf K}(a)^p}\equiv {\mathbf K}_p(a) $. This is the space of all the functions $\varphi(x)$ with support in $|x|\leq a$ and continuously differentiable up to order $p$. We have
\be 
{\mathbf K}_1(a)\supset {\mathbf K}_2(a)\supset\ldots \supset{\mathbf K}(a)\0
\ee

Another test function space is ${\mathbf S}$. It is the space of functions $\varphi(x)$ indefinitely differentiable in ${\mathbb R}$, that for $|x|\to\infty$ tend to 0 more rapidly than any power of $1/|x|$. This is also a countably normed space, the sequence of norms being defined by
\be
||\varphi||_p= {\rm sup}_{|k|,|q|\leq p}|x^k \varphi^{(q)}(x)|, \quad \quad p\in{\mathbb N}
\ee

A third example is the space   ${\mathbf Z}(a)$ of entire analytic functions in the complex variable $z=x+iy$, satisfying the inequalities
\be
|z^k \psi(z)|\leq C_k e^{a|y|}\0
\ee
where $C_k$ and $a$ are constants depending on $\psi$.
We  can define the sequence of norms
\be
||\psi||_p = {\rm sup}_{|k|\leq p} \,|z^k\psi(x+iy)| e^{-a|y|} \0
\ee
It follows that ${\mathbf Z}(a)$  is a countably normed space. Finally the space ${\mathbf Z}$  is the union of the spaces ${\mathbf Z}(a)$ (that is $a$ is generic).

All the above test function spaces can be generalized to many variables.
They are countably normed. A generic countably normed test function space will be denoted by the symbol $\boldsymbol \Phi$.

Any countably normed space is {\it metrizable}, that is one can define a metric $g(\varphi,\psi)$, which defines
the same topology as the countable set of norms. One can however prove that ${\mathbf K}(a)$ {\it is not a normed space}.

At this point one may wonder why we need a countably normed vector space as test function space: why is a simple normed space not enough? The answer is that the dual of a (possibly complete) normed vector space is itself a normed vector space. The latter cannot play the role of a space of distributions because in general we cannot associate a unique norm to distributions. The test function space must have a stronger topology than the normed one, in order to allow for the dual to accommodate distributions.

\subsection{Distributions}

Distributions are linear continuous functionals on a test function space ${\boldsymbol \Phi}$. The space of such functionals (dual space) will be denoted by ${\boldsymbol \Phi}'$. The rule for evaluating a functional $f$ over
a test function $\varphi$ will be denoted by $f(\varphi)=\langle f,\varphi\rangle$. For instance, for any
function $\varphi\in {\mathbf K}(a)$ we can define
\be 
\langle f,\varphi\rangle= \int_{-a}^a \varphi^{(m)}(x) d\mu(x)\0
\ee
where $m$ is a fixed positive integer and $\mu$ a function with bounded variation. One can prove that $f$ is a distribution. Since ${\boldsymbol \Phi}$ is countably normed we can define
\be
||f||_p = {\rm sup}_{||\varphi||_p\leq 1} \,|\langle f,\varphi\rangle|, \0
\ee
which is a norm in the dual space ${\boldsymbol \Phi}'_p$ of ${\boldsymbol \Phi}_p$. We have
\be
{\boldsymbol \Phi}'= \bigcup_{p=1}^\infty {\boldsymbol \Phi}_p'\0
\ee
that is, ${\boldsymbol \Phi}'$ is also countably normed and it is the union of an increasing sequence of
Banach spaces whose norm is weaker and weaker:
\be 
{\boldsymbol \Phi}'_1 \subset{\boldsymbol \Phi}'_2 \subset \ldots \subset  {\boldsymbol \Phi}'\0
\ee
The so-defined topology of the dual is called {\it strong}. There is also a {\it weak} topology. It is defined as follows.
Take a finite set of test functions $\varphi_1, \dots\varphi_m$. Then a neighborhood of 0 in ${\boldsymbol \Phi}'$
is defined by the $f'$s that satisfy
\be
|\langle f, \varphi_1\rangle|<\epsilon , |\langle f, \varphi_2\rangle|<\epsilon ,\ldots
|\langle f, \varphi_m\rangle|<\epsilon\0
\ee
We have the following definition:

A sequence of elements of ${\boldsymbol \Phi}'$, $f_n$, converge weakly to $f\in {\boldsymbol \Phi}'$ if and only if, for any $\varphi\in {\boldsymbol \Phi}$ we have
\be 
\lim_{n\to\infty} \langle f_n , \varphi\rangle = \langle f , \varphi\rangle\0
\ee 
It so happens that the dual ${\boldsymbol \Phi}'$ of a countably normed space ${\boldsymbol \Phi}$ is complete with respect to the weak topology.

Another definition is that of {\it perfect} space. A countably normed space is perfect if every bounded set is compact. One can prove that ${\mathbf K}(a)$ is perfect. If ${\boldsymbol \Phi}'$ is the dual of a perfect space the weak and strong topologies coincide.

\subsubsection{A representation theorem}

For all the spaces ${\mathbf K}(a)$,  ${\mathbf S}$ and  ${\mathbf Z}$ the following representation theorem holds

{\bf Representation theorem.} Any distribution  $f$ belonging to ${\boldsymbol \Phi}'$ admit the following representation
\be
\langle f, \varphi \rangle= \int dx\, f_0(x) P(D)\varphi(x)\label{repdistr}
\ee
where $f_0(x)$ is a locally summable function (Lebesgue integrable function) and $P(D)$ is a polynomial of $D=\frac d{dx}$.

This theorem can be extended to many variables.

A typical example is the Dirac delta function
\be
\langle \delta_a , \varphi\rangle = -\int dx \, \theta(x-a) \varphi'(x)\0
\ee
where $\theta$ is the step function.

\subsection{Examples}

A typical problem solved by distribution theory is the inversion of a polynomial $P(z)$ in the complex plane. The problem 
\be
P(z) f=1\label{inverseproblem}
\ee
has always a solution in terms of a distribution in ${\mathbf Z}'$. In particular the function $z$ or $z-z_0$ has an inverse represented by a distribution in ${\mathbf Z}'$. This result extends to a polynomial $P$ of many variables.
 
Let us consider a simpler case of inversion, as a paradigm for a later discussion, i.e. the inverse of $x$ on the positive real axis. To start with we define the distributions ${\rm x}^\lambda_+,{\rm x}^\lambda_-$ in ${\bf K}'(a)$ defined by the functions
\be 
x_+^\lambda= \left\{\begin{matrix}x^\lambda \quad\quad &x>0\\ 0 \quad\quad &x\leq 0\end{matrix}\right.,\quad\quad x_-^\lambda= \left\{\begin{matrix}0 \quad\quad &x\geq 0\\ x^\lambda \quad\quad &x<0\end{matrix}\right.\label{x+lafunct}
\ee 
where $\lambda$ is a complex parameter. For any test function $\varphi$, the rule
\be 
({\rm x}_+^\lambda, \varphi)= \int_0^\infty dx \, x^\lambda \,\varphi(x) \label{x+laint1}
\ee
is well defined for $\Re\lambda> -1$. We can extend it analytically via the formula
\be 
 \int_0^\infty dx \, x^\lambda \,\varphi(x)=\int_0^1 dx \, x^\lambda[\varphi(x)-\varphi(0)] +
\int_1^\infty dx\, x^\lambda \varphi(x) +\frac {\varphi(0)} {\lambda +1}\label{x+laint2} 
\ee
which is well defined for $\Re\lambda > -2$, $\lambda\neq -1$. We could continue and extend this formula for any $\lambda$, but for our needs we can stop here. Eq.(\ref{x+laint2}) defines a distribution ${\rm x}^\lambda_+$ for $\Re\lambda > -2$. It has a simple pole at $\lambda =-1$ with residue equal to $\delta(x)$. 

Two remarks. First, if $\varphi$ vanishes at the origin (\ref{x+laint2}) reduces to
$\int_0^\infty dx \, x^\lambda \,\varphi(x)$, which is well defined. Second, since for any test function $\varphi(x)$, $x \varphi(x)$ is still a test function, we can define 
$(x {\rm x}^\lambda_+, \varphi)=( {\rm x}^\lambda_+, x\varphi)$. Then eq.(\ref{x+laint2})
yields
\be 
({\rm x}_+^\lambda\, x, \varphi)= \int_0^\infty dx\,  \, x^{\lambda+1} \varphi(x)\label{x+laint3}
\ee

One can define in a similar way ${\rm x}^\lambda_-$ which extends to the negative real axis.
The distribution $|{\rm x}|^\lambda {\rm sgn\, x}= {\rm x}_+^\lambda-{\rm x}^\lambda_-$ is defined over the full real axis. Next one can define the distribution 
\be
({\rm x}+i0)^{-1}= {\rm x}^{-1} - i \pi \delta(x)\label{x+i0}
\ee
This is a particular case of the functional $({\rm x}+i0)^\lambda={\rm x}_+^\lambda + e^{i\lambda \pi} {\rm x}_-^\lambda$, which is entire in $\lambda$. In (\ref{x+i0})
${\rm x}^{-1}=\lim_{\lambda \to -1} ({\rm x}^\lambda_+-{\rm x}^\lambda_-)$ is simply given by
\be
({\rm x}^{-1}, \varphi)= \int_0^\infty dx \, x^{-1} (\varphi(x)-\varphi(-x))\label{x-1}
\ee
We will refer to this regularization as the {\it principal value} regularization. Of course
(\ref{x-1}) is well defined because $\varphi(x)-\varphi(-x)$ vanishes at the origin.

It is easy to see that $(x({\rm x}+i0)^{-1},\varphi)= (1,\varphi)$, i.e. $({\rm x}+i0)^{-1}$ can be interpreted as the inverse of $x$ on the full real axis. But
\be
({\rm x}-i0)^{-1}= {\rm x}^{-1} + i \pi \delta(x)\label{x-i0}
\ee
and ${\rm x}^{-1}$ itself have the same property. Thus the problem of defining a regularization of $\frac 1x$ on the full real axis does not have a unique solution. The various distributions that solve this problem differ from one another by a distribution with support on the singularity. 

It should be stressed that, if one chooses (\ref{x+i0}) or (\ref{x-i0}) as regularizations of $\frac 1x$, the delta function in the RHS is an integral part of the definition of the inverse, {\it not something to be added to the inverse itself}.

For later comparison let us quote also the distribution $|{\rm x}|^\lambda$ defined by
\be 
|{\rm x}|^\lambda= {\rm x}_+^\lambda+{\rm x}_-^\lambda\label{|x|}
\ee
which is analytic in $\lambda$ with a simple pole at $\lambda=-1$ with residue $\delta(x)$.
It is easy to see that $\lim_{\lambda\to -1} |{\rm x}|^\lambda$ is a well defined representative of $\frac 1{|x|}$.
\vskip 1cm

Let us pass next to a different kind of examples.
We recall that distributions can often be defined as limits of ordinary functions. For instance
\be
\delta(x) = \lim_{\e\to 0} \, \frac 1{\pi}\,\frac {\sqrt{\e}}{x^2+\e}\label{deltalim}
\ee
This follows from the fact that
\be 
\frac 1{\pi}\int_a^bdx \, \frac {\sqrt{\e}}{x^2+\e} = \frac 1{\pi} \left( \arctan \frac b{\sqrt{\e}} -
\arctan \frac a{\sqrt{\e}}\right),\label{deltalim2}
\ee
Thus whenever the interval $[a,b]$ includes 0, in the $\e\to 0$ limit we get 1, while if it does not contain 0 we get 0. This justifies (\ref{deltalim}), see \cite{Gelfand}. 

In the introduction we have quoted some empirical formulas used in SFT to regularize the inverse of $K$ or $K+\phi$. They are similar to the previous example. It is interesting to consider analogous formulas in the simpler but more rigorous context of distribution theory.

First let us notice that
\be 
\lim_{\e\to 0} \frac 1{|x|+\e} \, |x| = 1-\lim_{\e\to 0} \frac {\e}{|x|+\e} \label{1/|x|}
\ee
so that we can empirically interpret $\lim_{\e\to 0} \frac 1{|x|+\e}$ as the inverse of $|x|$, provided the second term in the RHS of (\ref{1/|x|}) vanishes (a similar manipulation leads to (\ref{first}) and (\ref{second})). We will proceed by  integrating it on any test function $\varphi\in {\bf K}(a)$. For simplicity we simulate
this by rectangular functions vanishing outside $|x|<a$. Therefore, up to a multiplicative constant, it is enough to integrate  between $-b$ and $c$ ($0<b,c<a$). The result is
\be
\int_{-b}^c dx\, \frac 1{|x|+\e}=\log\frac {(c+\e)(b+\e)}{\e^2} \approx -2\log(\e)+\ldots\0
\ee
Therefore the second piece on the RHS of (\ref{1/|x|}) vanishes, and $\lim_{\e\to 0} \frac 1{|x|+\e}$ is a formula for the inverse of $|x|$.

Proceeding in the same way for the inverse of $|x|^n+\e$  
\be
\lim_{\e\to 0} \frac 1{|x|^n+\e} \, |x|^n = 1- \lim_{\e\to 0} \frac {\e}{|x|^n+\e} \label{1/|x|n}
\ee
we find that
\be
\int_b^c dx\, \int_0^\infty dt \, e^{ -t(|x|^n+\e)}\sim\left\{\begin{matrix}  \e^{\frac {1-n}n}&\quad n>1\\ \e^{\frac {1-n}n}\log \e& \quad 0< n\leq 1\end{matrix}\right. \label{1/|x|n1}
\ee
and again the second term in the RHS of (\ref{1/|x|n}) vanishes, so we conclude that $\lim_{\e\to 0} \frac 1{|x|^n+\e} $ is a good empirical formula for the inverse of $|x|^n$.

The expression $\lim_{\e\to 0} \frac {\e}{|x|^n+\e}$ has support, if any, in $x=0$, but it is evident from the above that, for instance for $n>1$, we can have a nonvanishing, and finite, result (a delta-function-like object) only for $\lim_{\e\to 0} \frac {\e^{1 - 1/n} }{|x|^n+\e}$.

This is puzzling. In the following
we will meet an expression very similar to (\ref{deltalim}), but with an important difference.
The object we will have to discuss, (\ref{second}), is the analogue of
\be
\lim_{\e\to 0} \, \frac 1{\pi}\,\frac {\e}{x^2+\e}\label{deltalim3}
\ee
Due to the additional $\sqrt{\e}$ factor in the numerator  the anolog of
(\ref{deltalim2}) always vanishes in the $\e\to 0$ limit, even if the $[a,b]$ interval 
includes 0. Therefore
\be
\lim_{\e\to 0}\, \frac 1{\pi}\, \frac {\e}{x^2+\e}=0\label{deltalim4}
\ee
as a distribution. And it seems that, however we try to simulate (\ref{first}) and (\ref{second}), we always get 0.  Proceeding as in (\ref{1/|x|}) or (\ref{1/|x|n}) we can get a formula for the inverse of $|x|$ or $|x|^n$, but we cannot pick up the delta function contribution. The latter can be captured by introducing a complex parameter $\lambda$ and regularizing such expressions
as $x_\pm^\lambda$ and the like, as we have shown above.

\subsection{How large are the test function spaces?}

It would seem that there is an arbitrariness in the problem we have considered so far. Since a distribution $f$ is determined by its values on test functions, how do we know that this procedure uniquely determine $f$? In other words
suppose that for any $\varphi\in {\boldsymbol \Phi}$ we have $\langle f,\varphi\rangle=0$, is the space ${\boldsymbol \Phi}$ {\it rich } enough for us  to conclude that $f=0$?

This problem is well formulated by the following  {\it definition}: a test space ${\boldsymbol \Phi}$ is {\it rich enough} if, for any locally integrable function $f(x)$, existence of the integral
\be
\int dx \, f(x)\, \varphi (x) , \quad\quad \forall \varphi \in {\boldsymbol \Phi}\0
\ee
and
\be
\int dx \, f(x)\, \varphi (x) = 0, \quad\quad \forall \varphi \in {\boldsymbol \Phi}\0
\ee
imply that $f(x)=0$ almost everywhere. This means that $f$ is effectively zero if we restrict our consideration to elements of ${\boldsymbol \Phi}$ alone.

The spirit of this definition is to guarantee that the set of test functions is a powerful enough
filter that only very `fine' non-regular behaviors can pass through it. This filter cannot detect
for instance functions which are nonvanishing only in a set of measure zero, but it does detect any piecewise regular behaviour and, in particular, any regular behaviour.

It can be proved that all the test spaces considered above are rich enough. Once this condition is satisfied, we shall say that {\it a distribution is zero} if it vanishes when contracted with all the elements of $\boldsymbol{\Phi}$.

Since the duality rule can be formally extended to a space larger than the space of test functions, one can easily envisage a situation in which a zero distribution when evaluated on a non-test function does not vanish. But of course this is illegal.

\section{The dual of the Fock space}

We would like to consider now an analog of the rigged Hilbert space example presented in the introduction, based on a Fock space rather than on a function space. Let us consider the string oscillators $\alpha_n$, with the algebra $[\alpha_m,\alpha_n]= m \delta_{n+m,0}$. We construct the corresponding Fock space $\boldsymbol {\cal F}$ by acting on the vacuum $|0\rangle$ with the creation operators $\alpha_n$ with $n<0$.  $\boldsymbol {\cal F}$ is the linear span of all the states of the type 
\be
|\phi_{n_1,n_2,\ldots,n_s}\rangle=\alpha_{-n_1} \alpha_{-n_2} \ldots \alpha_{-n_s} |0\rangle, \quad\quad n_i\in {\mathbb Z}_+\label{phistates}
\ee
As is well known the oscillator algebra defines a scalar product once we assume that $\langle 0|0\rangle=1$. This also implies the definition of a norm $||\phi||= \sqrt{\langle \phi|\phi\rangle}$. The completion of  $\boldsymbol {\cal F}$ with respect to this norm 
gives rise to a Hilbert space ${\boldsymbol{\EH}}\equiv\overline{\boldsymbol {\EF}}$ which contains $\boldsymbol {\cal F}$ as a dense subset. 

Like in the quantum mechanical example in the introduction, however, this Hilbert space does not contain all the interesting states in SFT. For instance, it does not contain wedge states and, generically, surface states, because such states do not have a finite norm. Similar to the rigged Hilbert space we would like to find a space larger than ${\boldsymbol{\EH}}$. To this end we have to give up the scalar product topology in $\boldsymbol {\cal F}$ and  introduce a stronger one.
The linear functionals which are continuous with respect to this stronger topology form the dual 
space $\boldsymbol {\cal F}'$.  If suitably chosen\footnote{To our best knowledge this exercise has not yet been carried out in the literature.} the space $\boldsymbol {\cal F}'$ will allow us to embed surface states and the like  and to study limits such as (\ref{first}).

The Fock space $\boldsymbol {\cal F}$ plays the role of  $\boldsymbol {\cal R}$, the space of regular objects. Therefore it is natural to say that, if a generic state vanishes when contracted with all Fock space states, it is zero. 

The limitation to Fock space states is essential, when using this criterion, otherwise it is easy to construct a counterexample. Consider a definite 
Fock space state $\phi_0$. It obviously has finite contractions with all the other Fock states.
Therefore if we consider $\phi_{\epsilon}= \epsilon \phi_0$, we have that 
$\lim_{\epsilon\to 0}\phi_{\epsilon}$ vanishes when contracted with all the states of the Fock space. Thus $\lim_{\epsilon \to 0} \phi_{\epsilon}=0$ according to the above criterion.
If, on the other hand, we contract $\phi_\epsilon$ with states like $\frac 1{\epsilon} \phi$, where $\phi$ is any Fock space state, in the limit $\epsilon\to 0$ we find a finite result.
This is not surprising since the norm of $\frac 1{\epsilon} \phi$ becomes infinite in the  
 $\epsilon\to 0$ limit.

\section{Origin of the problem}

Let us recall the problem that we want to formalize in terms of distribution theory and highlight
the main points needed in the subsequent discussion.
 
In \cite{BMT}, the $K,B,c$ algebra
\be
K=\frac\pi2K_1^L| I\rangle, \quad \quad
B=\frac\pi2B_1^L| I\rangle,\quad\quad c= c\left(\frac12\right)|I\rangle,
\label{KBc}
\ee
was enlarged as follows. In the sliver frame (obtained by mapping the UHP to
an infinite cylinder $C_2$ of circumference 2, by the sliver map
$\tilde z=\frac2\pi\arctan z$), we consider a (relevant) matter operator
\be \phi=\phi\left(\frac12\right)|I\rangle\label{phi} \ee with the
properties \be \,[c,\phi]=0,\quad\quad \,[B,\phi]=0,\quad\quad
\,[K,\phi]= \del\phi,\label{proper}
\ee
In this new algebra $Q$ has the following action:
\be
Q\phi=c\del\phi+\del c\delta\phi.
\label{actionQ}
\ee
It can be easily proven that
\be
\psi_{\phi}=c\phi-\frac 1{K+\phi}(\phi-\delta\phi) Bc\del c\label{psiphi}
\ee
does indeed satisfy the OSFT equation of motion
\be Q\psi_{\phi}+\psi_{\phi}\psi_{\phi}=0\label{eom}.
\ee
It is clear
that (\ref{psiphi}) is a deformation of the Erler--Schnabl solution,
see \cite{ErlerSchnabl}, which can be recovered for $\phi=1$.

It was pointed out in \cite{BMT,BGT1,BGT3} that a new nontrivial solution requires the following conditions to be satisfied:
\begin{itemize}
\item (a) $\frac 1{K+\phi}$ is singular, but
\item (b) $\frac 1{K+\phi}(\phi-\delta\phi)$ is regular and
\item (c) $\frac 1{K+\phi}(K+\phi)=1.$
\end{itemize}

These conditions seem at first contradictory. But the reader is advised to think in terms of the following simple paradigm: in the complex $z$-plane $\frac 1z$ is singular, but ${\frac 1z}\cdot z=1$ everywhere, by continuity. In this light the above conditions do not look anymore so puzzling. It is the purpose of this paper to clarify that, once they are appropriately specified, they do not contain any contradiction.

To parametrize the RG flow in 2d, let us introduce a variable $u$,  where
$u=0$ represents the UV and $u=\infty$ the IR, and rewrite $\phi$ as
$\phi_u$, with $\phi_{u=0}=0$. Then we require for $\phi_u$ the
following properties under the coordinate rescaling $f_t(z)=\frac zt$
\be
f_t\circ\phi_u(z)=\frac 1t\,\phi_{tu}\left(\frac
zt\right)\label{cnd1}
\ee
and, most important, that the partition function
\be
g(u)\equiv Tr[e^{-(K+\phi_u)}]=\left\langle
e^{-\int_0^1ds\, \phi_{u}(s) }\right\rangle_{C_1},\label{g(u)Tr}
\ee
satisfies the asymptotic finiteness condition
\be
\lim_{u\to\infty}\left\langle e^{-\int_0^1ds\, \phi_u(s)
}\right\rangle_{C_1}={\rm finite}.\label{cnd3}
\ee
It was pointed out in \cite{BMT} that this satisfies the first two conditions above i.e.
guarantees not only the regularity of the solution but also its
`non-triviality', in the sense that if this condition is satisfied,
it cannot fall in the same class as the ES tachyon vacuum solution.
It would seem that the last condition above cannot be satisfied in
view of the first. But this is not the case. This is precisely the issue that has to be formalized in terms of distribution theory, see below. 

To continue we consider in the sequel a specific relevant operator $\phi_u$
and the corresponding SFT solution. This operator generates an exact
RG flow studied by Witten in \cite{Witten2}{},  see also
\cite{Kutasov}{}, and is based on the operator (defined in the
cylinder $C_T$ of width $T$ in the arctan frame)
\be
\phi_u(s) = u(X^2(s)+2\ln u +2 A)\label{TuCT1},
\ee
where $A$ is a constant first
introduced in \cite{Ellwood}.  This implies that on the unit disk $D$ we have,
\be \phi_u(\theta) = u (X^2(\theta)+ 2\ln \frac {Tu}{2\pi} +2
A).\label{TuDb1}
\ee

Setting
\be
g_{A}(u)= \langle e^{-\int_0^1ds \,
\phi_u(s)}\rangle_{C_1} \label{gAu1} \ee we have \be g_{A}(u)
=\langle e^{- \frac 1{2\pi } \int_{0}^{2{\pi}} d\theta \, u\Bigl{(}
X^2(\theta) + 2 \ln \frac u{2\pi}+2 A\Bigr{)}}\rangle_{D}.  \0
\ee
According to \cite{Witten2},
\be
g_{A}(u) = Z(2u)e^{-2u (\ln \frac
u{2\pi}+A)},\label{gAu2} \ee where \be Z(u)= \frac 1{\sqrt{2\pi}}
\sqrt{u} \Gamma(u)e^{\gamma u}\label{Z(u)}
\ee

Finiteness for $u\to\infty$ requires $A= \gamma -1+\ln
4\pi$, which implies
\be
g_{A}(u)\equiv g(u)= \frac 1{\sqrt{2\pi}}
\sqrt{2u} \Gamma(2u) e^{2u(1-\ln (2u))}\label{gA1unorm}
\ee
and
\be
\lim_{u\to\infty} g(u) = 1.\label{inflimitgA1u} \ee
 Therefore the $\phi_u$ just introduced satisfies all the required properties and consequently
$\psi_u\equiv \psi_{\phi_u}$ must represent a D24 brane solution.

In \cite{BMT} the expression for the energy of the lump solution was determined by evaluating a three--point
function on the cylinder $C_T$ of circumference $T$ in the arctan frame. It is given by
\be
E[\psi_u]&=&-\frac16 \langle \psi_u \psi_u\psi_u\rangle\0\\
&=&\frac16 \int_0^\infty d(2uT)\; (2uT)^2\int_0^1
dy\int_0^{y} dx\,\frac4\pi \,\sin\pi x\,\sin\pi y\,\sin\pi(x-y)\label{Efinal}\\
&&\cdot g(uT)\Bigg\{-\Big(\frac{\partial_{2uT}g(uT)}{g(uT)}\Big)^3
+G_{2uT}(2\pi x)G_{2uT}(2\pi(x-y))G_{2uT}(2\pi y)\0\\
&&-\frac 12 \Big(\frac{\partial_{2uT}g(uT)}{g(uT)}\Big)\Big(G_{2uT}^2(2\pi
x)+G_{2uT}^2(2\pi(x-y))+G_{2uT}^2(2\pi y)\Big) \Bigg\}.\0
\ee
where $G_u(\theta)$ represents the
correlator on the boundary, first determined by Witten, \cite{Witten2}:
\be
G_u(\theta)= \frac {1}{u} +2 \sum_{k=1}^\infty
\frac {\cos (k\theta)}{k+u} \label{Gutheta}
\ee
Moreover ${\cal E}_0(t_1,t_2,t_3)$ represents the ghost three--point function in $C_T$.
\be
{\cal E}_0(t_1,t_2,t_3)=\left\langle Bc\del c(t_1+t_2)\del c(t_1) \del c(0)
\right\rangle_{C_T} = -\frac 4{\pi} \sin \frac {\pi t_1}T \sin \frac
{\pi(t_1+t_2)}T \sin \frac{\pi t_2}T.\label{E0}
\ee
Finally, to get (\ref{Efinal}) a change of variables $(t_1,t_2,t_3)\to (T,x,y)$, where \be x=\frac{t_2}{T},
\quad\quad y=1-\frac{t_1}{T}.\0
\ee
is needed.

The expression (\ref{Efinal}) was evaluated in \cite{BGT1}. As it turns out, this expression has 
a UV ($s\approx 0$, setting $s=2uT$) singularity, which must be subtracted away. Therefore the result one
obtains in general will depend on this subtraction. In \cite{BGT1} it has been pointed out that {\it a physical
significance can be assigned only to a subtraction-independent quantity}, and it has been shown how to define and evaluate such a quantity. First a new solution to the EOM, depending on a parameter $\epsilon$, was introduced  
\be
\psi_u^{\epsilon}= c(\phi_u+\epsilon) - \frac 1{K+\phi_u+\epsilon} (\phi_u+\epsilon -\delta \phi_u) Bc\partial c.
\label{psiphieps}
\ee
its energy being 0 (after a similar UV subtraction as in the previous case) in the $\epsilon\to 0$ limit.
Then, using it, a solution to the EOM at the tachyon condensation vacuum was obtained. The equation of motion at the tachyon vacuum is
\begin{align}\label{EOMTV}
{\cal Q}\Phi+\Phi\Phi=0,\quad {\rm where}~~{\cal Q}\Phi=Q\Phi+\psi_u^\epsilon\Phi+\Phi\psi_u^\epsilon.
\end{align}
One can easily show that
\begin{align}\label{psiupsie}
\Phi_0^\epsilon=\psi_u-\psi_u^\epsilon
\end{align}
is a solution to (\ref{EOMTV}). The action at the tachyon vacuum is $-\frac12\langle{\cal
Q}\Phi,\Phi\rangle-\frac13\langle\Phi,\Phi\Phi\rangle.$ Thus the energy of of the lump, $E[\Phi_0]$, is
\be\label{TVlumpenergy}
E[\Phi_0]&=&-\lim_{\epsilon\to 0}\frac16\langle\Phi_0^\epsilon,\Phi_0^\epsilon\Phi_0^\epsilon\rangle\0\\
&=&
-\frac16\lim_{\epsilon\to 0} \big[\langle\psi_u,\psi_u\psi_u\rangle-\langle\psi_u^\epsilon,\psi_u^\epsilon\psi_u^\epsilon\rangle
-3\langle\psi_u^\epsilon,\psi_u\psi_u\rangle+3\langle\psi_u,\psi_u^\epsilon\psi_u^\epsilon\rangle\big].
\ee
The integrals in the four correlators on the RHS, are IR ($  s\to\infty$) convergent.
The UV subtractions necessary for each correlator are always the same, therefore they cancel out. In \cite{BGT1}, after UV subtraction, we obtained
\be
 &&-\frac 16 \langle\psi_u,\psi_u\psi_u\rangle=\alpha+\beta, \quad\quad \lim_{\epsilon\to 0}\langle\psi_u^\epsilon,\psi_u^\epsilon\psi_u^\epsilon\rangle=0\0\\
&&\frac 16 \lim_{\epsilon\to 0}\langle\psi_u^\epsilon,\psi_u\psi_u\rangle=\alpha-\frac 23 \beta,
\quad\quad   \frac 16 \lim_{\epsilon\to 0} \langle\psi_u,\psi_u^\epsilon\psi_u^\epsilon\rangle=
\alpha -\frac 13 \beta\label{finalBGT1}
\ee
where $\alpha+\beta\approx 0.068925$ was evaluated numerically and $\alpha= \frac 1{2\pi^2}$ was calculated analytically. So $E[\Phi_0]= \alpha$ turns out to be precisely the D24-brane energy. In \cite{BGT2} the same result was extended to any Dp-brane lump.

The problem that needs to be formalized in terms of distribution theory is related to condition 3 above, i.e. the existence of the inverse of $K+\phi_u$.

\section{The space of test string fields}

\subsection{Preliminary discussion}

The problem we would like to discuss here is the existence of the inverse of $K+\phi_u$. As we have pointed out $K+\phi= (K_1^L+\phi_u(\frac12))|I\rangle$, where $K_1^L$ is the left translation operator, a symmetric operator in the Fock space, and $|I\rangle$ is the star algebra identity. The spectrum of $\EK_u\equiv K_1^L+\phi_u(\frac12)$, which is also a symmetric operator, lies in the real axis and is likely to include also the origin. If it does and the identity string field contains the zero mode of $\EK_u$, then a problem of invertibility arises. Proceeding empirically, the obstruction to invert  $K+\phi_u$ is measured by the expression 
\be\label{A}
{\cal A}_0= \lim_{\e\to 0}{\cal A}_{\e}, \quad\quad {\cal A}_{\e}=\frac {\e}{K+\phi_u+\e}
\ee
since
\be
\frac 1{K+\phi_u+\e} (K+\phi_u)=1- {\cal A}_{\e}\label{Ainverseobstr}
\ee

This quantity, whatever it is, is nonvanishing only where $K+\phi_u$ vanishes, i.e. in correspondence with the zero mode of $\EK_u $. ${\cal A}_{0}$ has support, if any, only on this zero mode. It is a {\it distribution-like object} and must be treated within the formalism of distribution theory. Of course  the latter must be suitably generalized to the framework of SFT in which a position in space (for instance $r=0$ in ordinary field theory) is replaced by a string configuration (for instance the state representing the zero mode above). The correct evaluation of ${\cal A}_\e$ is of upmost importance, for a naive manipulation of the equation of motion leads to
\be 
Q\psi_u+\psi_u\psi_u=  \lim_{\e\to 0}\frac {\e}{K+\phi_u+\e}(\phi_u-\delta \phi_u)c\partial c
\label{eomviolation}
\ee
i.e. to an apparent violation of the equation of motion, \cite{EM}.

Our purpose in the sequel is to set up a mathematical framework in order to be able to unambiguously interpret such expressions as the RHS of (\ref{eomviolation}) and decide whether such a violation is apparent or real. As we have seen in the previous section 
in function theory, distributions are objects of the dual of a topological vector space. But this topological vector space is not a generic one, rather it is a space of functions with certain properties, so that its dual can be regarded as a space of `derivatives' of locally integrable functions. Our aim is to investigate on the possibility to interpret the inverse of $K+\phi_u$ (and ${\cal A}_{\e}$) as a distribution, i.e. a functional in a suitable topological vector space. 

The first step consists in constructing this topological vector space. It is probably true that one should start from a very general and abstract point of view like the one envisaged by L.Schwartz, \cite{Schwartz}. Here we take a more modest and unsophisticated, but constructive, attitude, by using our knowledge of correlators in open SFT. To be treatable, the vector space should have properties that makes it similar to a space of functions, and the duality rule (i.e the rule by which we can evaluate a functional over the test states) should preferably be represented by an integral. This would allow us to use the analogy with ordinary distribution theory as close as possible. Fortunately this is possible in the present case, thanks to the Schwinger representation of the inverse of  $K+\phi_u$:
\be
\frac 1{K+\phi_u}= \int_0^\infty dt\, e^{- t(K+\phi_u)}\label{Schw}
\ee
This representation makes concrete the abstract properties of the functional in question and `localizes' the zero mode of $\EK_u$ at $t=\infty$ (for the representation (\ref{Schw}) becomes singular when $K+\phi_u$ vanishes). This `localization property' makes our life much easier because it allows us to formulate the problem of defining test states, dual functionals and their properties in terms of their $t$ dependence via the Schwinger representation (\ref{Schw}).

We anticipate that the test states cannot be 
`naked' Fock space states because there is no way to interpret such states {\it as good test states} (see the discussion in the next section).

\subsection{Good test string fields}

Let us now construct a set of string states that have good properties in view of forming the
topological vector space of test states we need for our problem\footnote{This section is based on the results of \cite{BGT3}}. Our final {\it distribution} or {\it regularization} will be analogous to the principal value regularization of ${\rm x}^{-1}$ in section 2.3, which is characterized by  ${\rm x}^{-1}$ being evaluated on test functions vanishing at the origin.

First of all the states we are looking for must be such that the resulting contractions with $\Gamma(\e)= {\cal A}_{\e} (\phi_u-\delta\phi_u) c \partial c$ be nonsingular (with respect to singularities due to collapsing points). But, especially, they must be characterized by integrable behaviour in the UV and, ignoring the overall $e^{-\e t}$ factor, in the IR. The IR corresponds to $t\to\infty$, where, as was noticed above, the zero mode of $\EK_u$ is `localized'. Therefore the IR behaviour will be crucial in our discussion.  

Consider states created by multiple products of the factor $H(\phi_u,\e)=\frac 1{K+\phi_u+\e} (\phi_u-\delta \phi_u)$ and contract them with 
\be
\Lambda(\e)= \frac 1{K+\phi_u+\e} (\phi_u-\delta\phi_u) c \partial c\label{Lambda}
\ee 
From what we said above we are looking for contractions which are finite and whose $\e\to 0$ limit is continuous. More precisely, let us define
\be
{\mathbf \Psi}_n(\phi_u,\e)= H(\phi_u,\e)^{n-1}   Bc\partial c\, H(\phi_u,\e),\quad\quad n\geq 2\label{Psin}
\ee
Contracting with $\Lambda(\e)$: $\langle {\mathbf \Psi}_n(\phi_u,\e)Bc\partial c B\,\Lambda(\e)\rangle$, we obtain a correlator
whose IR and UV behaviour (before the the $\e\to 0$ limit is taken) is not hard to guess. The correlators take the form 
\be
&&\langle {\mathbf \Psi}_n(\phi_u,\e)Bc\partial c B\,\Lambda(\e)\rangle=
\int_0^\infty ds\, s^n e^{-\tilde \eta s}\,g(s) \cdot\label{Hn+1}\\
&&\quad\quad\quad\cdot\int \prod_{i=1}^ndx_i\,\EE \left( \left(-\frac {\del g(s)}{g(s)}\right)^{n+1} + \ldots+ \left(-\frac {\del g(s)}{g(s)}\right)^{n-k+1} G_s^k +\ldots+ G_s^{n+1}\right)\0
\ee
where the notation is the same as in the previous section ($s=2uT$ and $\tilde \eta =\frac {\e}{2u}$), but we have tried to make it as compact as possible.
The angular variables $x_i$ have been dropped in $\EE$ and $G_s$. Using the explicit form of $G_s$, \cite{BMT}, expanding the latter with the binomial formula and integrating over the angular variables, one gets
\be 
\int \prod_{i=1}^ndx_i\,\EE\,G_s^k = \sum_{l=0}^k \frac 1{s^{k-l}} \, \sum_{n_1,\ldots,n_{l}}\,
\frac {P_l(n_1,\ldots,n_{l})}{Q_l(n_1,\ldots,n_{l})} \prod_{i=1}^l \frac 1{p_i(n_1,\ldots,n_{l})+s}\label{intGs}
\ee
The label $l$ counts the number of cosine factors in each term.
Here $n_i$ are positive integral labels which come from the discrete summation in $G_s$; $p_i(n_1,\ldots,n_{l})$ are polynomials linear in $n_i$. Next, $P_l$ and $Q_l$ are polynomials in $n_i$
which come from the integration in the angular variables. Every integration in $x_i$ increases by 1 the difference in the degree of $Q_l$ and $P_l$, so that generically ${\rm deg} Q_l-{\rm deg} P_l= n$. But in some subcases the integration over angular variables gives rise to Kronecker deltas among the indices, which may reduce the degree of $Q_l$. So actually the relation valid in all cases is  ${\rm deg} Q_l\geq {\rm deg} P_l$. But one has to take into account that the number of angular variables to be summed over decreases accordingly. 

We are now in the condition to analyze the UV behaviour of (\ref{Hn+1}).
Let us consider, for instance, the first piece  
\be 
\sim \int_0^\infty ds\,e^{-\tilde \eta s} s^{n}\,g(\frac s2) \left( \frac {\del_s g(\frac s2)}{g(\frac s2)}\right)^{n+1}\label{piece}
\ee
Since in the UV $g(\frac s2) \approx \frac 1{\sqrt s}$, it is easy to see that the UV behaviour of the overall integrand is $\sim s^{-\frac 32}$, independently of $n$. As for the other terms, let us consider in the RHS of (\ref{intGs}) the factor that multiplies $\frac 1{s^{k-l}}$ (for $l\geq 2$). Setting $s=0$, the summation
over $n_1,\ldots,n_{l-1}$ is always convergent, so that the UV behaviour of each term in the summation is given by the factor $\frac 1{s^{k-l}}$, with $2\leq l\leq k$. It follows that the most UV divergent term corresponds to $l=0$, i.e. $\sim \frac 1{s^k}$. Since in (\ref{Hn+1}) this is multiplied by 
\be
s^n \,g(\frac s2 )\left(-\frac {\del g(\frac s2)}{g(\frac s2)}\right)^{n-k+1}\label{rest}
\ee
we see that the UV behaviour of the generic term in (\ref{Hn+1}) is at most as singular as $\sim s^{-\frac 32}$.
{\it In conclusion the states ${\mathbf \Psi}_n$, when contracted with $\Lambda_\e$, give rise to the same kind of UV singularity $\sim s^{-\frac 32}$}. Now, for any two such  states, say
${\mathbf \Psi}_n$ and  $ {\mathbf \Psi}_{n'}$, we can form a suitable combination such that the UV singularity cancels. In this way we generate infinite many states, say ${\mathbf \Phi}_n$, which, when contracted
with $\Lambda(\e)$, give rise to UV convergent correlators. 

Let us consider next the IR properties ($s\gg 1$). All the correlators contain the factor $e^{-\tilde \eta s}$ which renders them IR convergent. So as long as this factor is present the states ${\mathbf \Phi}_n$ are both IR and UV convergent when contracted with $\Lambda(\e)$.
But the crucial, \cite{BGT3}, IR properties (in the limit $\e\to 0$) are obtained by ignoring the exponential factor. This is in order to guarantee the continuity of the $\e\to 0$ limit. So, in analyzing the IR properties we will ignore this factor. The first term (\ref{piece}) is very strongly convergent in the IR, because $\del_s g(\frac s2)\approx \frac 1{s^2}$, while $g(\frac s2) \to 1$. For the remaining terms let us consider in the RHS of (\ref{intGs}) the factor that multiplies $\frac 1{s^{k-l}}$ (for $l\geq 2$). To estimate the IR behaviour it is very important to know the degree difference between the polynomials $Q_l$ and $P_l$. Above we said that this difference is always nonnegative. In principle it could vanish, but from the example with   $n=2$, see \cite{BGT1}, we know that there are cancellations and that in fact the difference in degree is at least 2.
If this is so in general, we can conclude that the IR behaviour of the summation in the RHS of (\ref{intGs})
with fixed $l$ is $\sim \frac 1{s^l}$. However, in order to prove such cancellations, one would have to do detailed calculations, which we wish to avoid here. So we will take the pessimistic point of view and assume that, at least for some of the terms, ${\rm deg} Q_l={\rm deg} P_l$ (in which case there remains only one angular integration). In this case the IR behaviour of the corresponding term cannot decrease faster than $\sim \frac 1{s^{l-1}}$. This has to be multiplied by $\sim \frac 1{s^{k-l}}$ and by the IR behaviour of (\ref{rest}). This means that the least convergent term
with fixed $k$ in(\ref{intGs}) behaves as $\sim \frac 1{s^{n-k+1}}$. Since $k\leq n+1$, we see that in the worst
hypothesis in the integral (\ref{Hn+1}) there can be linearly divergent terms. If this is so the UV converging ${\mathbf \Phi}_n$ states are not good test states. However we can repeat for the IR singularities what we have done for the UV ones. Taking suitable differences of the ${\mathbf \Phi}_n$'s (this requires a two steps process, first for the linear and then for the logarithmic IR singularities\footnote{In the, so far not met, case where a $\log s$ asymptotic contribution appears in the integrand one would need a three step subtraction process.}), we can create an infinite set of states, ${\mathbf \Omega}_n$, which, when
contracted with $\Lambda(\e)$, yield a finite result and whose $\e\to 0$ limit is continuous. Upon applying $\Gamma(\e)$, instead of $\Lambda(\e)$, such contractions of course vanish. 
These ${\mathbf \Omega}_n$ are therefore good (and nontrivial) test states. They are annihilated by $\Gamma(\e)$.  

We remark that in eq.(\ref{Psin}) the presence of $\e$ in $H(\phi_u,\e)$ is not essential, because in estimating the IR behaviour we have not counted the $e^{-\tilde \eta s}$ factor. Using $\frac 1{K+\phi_u}$ everywhere instead of $\frac 1{K+\phi_u+\e}$, would lead to the same results. This means that
contracting the $\Omega_n$ states with $\Lambda_\e$ leads to finite correlators {\it with or without $\e$}. We stress again that {\it the $\e\to 0$ limit of such correlators is continuous}. This is the real distinctive features of good test states. The property of annihilating $\Gamma(\e)$, is a consequence thereof. This remark will be used later on.

The $\Omega_n(\phi_u,\e)$ are however only a first set of good test states. One can envisage a manifold of other such states. Let us briefly describe them, without going into too many details. For instance, let us start
again from (\ref{Psin}) and replace the first $H(\phi_u,\e)$ factor with $\frac 1{K+\phi_u+\e} uX^{2k}$ (the term $\delta \phi$ can be dropped). In this way we obtain a new state depending on a new integral label $k$.
However replacing $X^2$ with $X^{2k}$ is too rough an operation, which renders the calculations unwieldy, because it breaks the covariance with respect to the rescaling $z \to \frac zt$. It is rather easy to remedy by studying the conformal transformation of $X^{2k}$. The following corrected replacements will do:
\be
uX^2 &\to& u\left(X^2 +2 (\log u+\gamma)\right)=\phi_u\equiv\phi_u^{(1)} \0\\
uX^4 &\to&u\left( X^4 +12 (\log u+\gamma) X^2+ 12 (\log u+\gamma)^2\right)\equiv \phi_u^{(2)}\0\\
&\ldots&\0\\
uX^{2k} &\to&u\left( \sum_{i=0}^k \frac {(2k)!}{(2k-2i)! i!} \left(\log u+\gamma\right)^i X^{2k-2i}\right)\equiv \phi_u^{(k)}\label{X2k}
\ee
The role of the additional pieces on the RHS is to allow us to reconstruct the derivatives of $g(s)$ in computing the correlators, as was done in \cite{BMT}. 

Now let us denote by ${\mathbf \Psi}_n^{(k)}$ the $n$-th state (\ref{Psin}) where $\phi_u-\delta\phi_u$ in the first $H(\phi_u,\e)$ factor is replaced by $\phi_u^{(k)}$. Contracting it with $\Lambda(\e)$ it is not hard to see
that the term (\ref{piece}) will be replaced by
\be 
\sim \int_0^\infty ds\,e^{-\tilde \eta s} \, s^{n}\,g(\frac s2) \left( \frac {\del_s g(\frac s2)}{g(\frac s2)}\right)^{n+k}\label{piecenew}
\ee
with analogous generalizations for the other terms. It is evident from (\ref{piecenew}) that the UV behaviour becomes more singular with respect to (\ref{piece}) while the IR one becomes more convergent. This is a general 
property of all the terms in the correlator. Thus fixing $k$ we will have a definite UV singularity, the same up
to a multiplicative factor for all ${\mathbf \Psi}_n^{(k)}$. Therefore by combining a finite number of them we can eliminate the UV singularity and obtain another infinite set of UV convergent states ${\mathbf \Omega}_n^{(k)}$ for any $k$ (${\mathbf \Omega}_n^{(1)}$ will coincide with the previously introduced ${\mathbf \Omega}_n$). In general they will be IR convergent (IR subtractions may be necessary for $k=2$ beside $k=1$).

It goes without saying that the previous construction can be further generalized by replacing in (\ref{Psin}) more than one $X^2$ factors with higher powers $X^{2k}$.

Qualitatively one can say that the correlators discussed so far have the form of an $s$ integral 
\be 
\int_0^\infty ds \, F(s)\label{qualitative}
\ee
where the $F(s)$ at the origin behaves as $s^{\frac k2}$, with integer $k\geq -1$, and $F$ with all possible $k$'s are present. At infinity, excluding the $e^{-\tilde \eta s}$, $F(s)$ behaves
as $\frac 1{s^p}$, for any integer $p\geq 2$. In addition, at infinity, we have any possible exponentially decreasing behaviour. 

It is evident that {\it all} the states ${\mathbf \Omega}^{(k)}_n$ annihilate $\Gamma(\e)$.
In fact the  ${\mathbf \Omega}^{(k)}_n$ are analogous of the test functions $\varphi(x)$ that vanish at the origin, like the one considered in section 2.3 for the regularization of the distribution ${\rm x}^{-1}$. On the other hand, the only possibility of getting a nonzero result while contracting $\Gamma(\e)$ with test states is linked, as usual, to correlators corresponding to IR linearly divergent integrals (without the exponential $e^{-\e t}$). Now, such integrals are characterized by the fact that their $\e\to 0$ limit is discontinuous, therefore the corresponding states can hardly be considered {\it good test states}. 
The true question we have to ask, then, is whether the good test states we have constructed are `enough'.

\section{The topological vector space of test states }

Above we have introduced an infinite set of good test states which will be denoted generically
by ${\mathbf \Omega}_{\alpha}$, $\alpha\in {\bf A}$ being a multi-index. We recall that in ${\mathbf \Omega}_{\alpha}$ there is also a dependence on the parameter $\e$. Such a dependence improves the IR convergence properties. 
The linear span of these state will be denoted by ${\boldsymbol {\EF}}$.
It is a vector space. 
The problem now is to define a topology on it. First of all we define for any two states $\Omega_\alpha, \Omega_\beta$ 
\be
\langle\Omega_\alpha|\Omega_\beta \rangle \equiv \langle {\mathbf \Omega_\alpha} B c\partial c B \Lambda_\e\rangle  \langle {\mathbf \Omega_\beta} B c\partial c B \Lambda_\e\rangle \, f(\alpha,\beta) \label{innerproduct}
\ee
where in the RHS feature the previously defined correlators, and $f(\alpha,\beta)$ is a generic real symmetric function of $\alpha$ and $\beta$\footnote{The numerical factor $f(\alpha,\beta)$ is needed in order to avoid a catastrophic degeneracy of the inner product. Provided it is sufficiently generic its actual expression is not important in the sequel.}. From the analysis of the previous subsection this is a finite number, generically nonvanishing. Whenever a correlator of this kind depends on $\e$, the limit $\e\to 0$ exists and is finite. The definition (\ref{innerproduct}) can be extended by linearity to all finite combinations of the vectors ${\mathbf \Omega}$ (we stress that linearity is imposed by hand).
Thus ${\boldsymbol {\EF}}$ is an inner product space. This inner product is not a scalar product in general. We will assume that it is nondegenerate (i.e. there are no elements with vanishing inner product with all the elements of the space)\footnote{If the inner product is degenerate the subsequent construction can be equally carried out, but it is more complicated, see for instance \cite{Bognar}.} . The existence of an inner product does not mean by itself that ${\boldsymbol {\EF}}$ is a topological vector space, but it is possible to utilize it to define a topology. 

\subsection{Seminorm topology}

There are various ways to introduce a
topology in an inner product space ${\cal V}$. We will use seminorms. 
Let us denote by  $x,y,...$ the elements of ${\cal V}$, and by $(x,y)$ the inner product.
A {\it seminorm} is a function in 
${\cal V}$   that satisfies the following axioms
\be 
&& p(x)\geq 0\0\\
&& p(a x) =|a| p(x),\quad a\in{\mathbb C}\label{seminorm}\\
&& p(x+y)\leq p(x)+ p(y) \0
\ee
Once we have an (infinite) family $p_\gamma$ ($\gamma$ is a generic index) of seminorms we can define a topology $\tau$ 
in the following way: a subset $V$ is open if for any $x\in V$ there is a finite subset
$p_{\gamma_1},p_{\gamma_2},\ldots, p_{\gamma_n}$ of seminorms and a positive number $\epsilon$, such that, any other element $y$ satisfying $p_{\gamma_j}(x-y)<\epsilon$, for $j=1,\ldots n$ belongs to $V$.
A topology $\tau$ is {\it locally convex} if the vector space operations are continuous in $\tau$ and if a $\tau$-neighborhood of any point $x$ contains a convex neighborhood of the same point.

What we wish is of course a topology strictly related to the inner product. Therefore we introduce
the concept of {\it partial majorant}. A partial majorant of the inner product $(\cdot,\cdot)$ is
a topology $\tau$ which is locally convex and such that for any $y\in {\cal V}$ the function
$\varphi_y(x)=(x,y)$ is $\tau$-continuous. If  $(x,y)$ is jointly $\tau$-continuous we say that $\tau$ is a {\it majorant}.  Apart from being locally convex (the minimal requirement for the validity of the Hahn-Banach theorem), the main endowment of a majorant topology is the continuity of the inner product simultaneously in both entries. 
 
In addition we say that a topology $\tau$ is {\it admissible} if 1) $\tau$ is a partial majorant and 2) for any linear $\tau$-continuous functional $\varphi_0(x)$ there is an element $y_0\in{\cal V}$ such that $\varphi_0(x)=(x,y_0)$. That is, all the continuous linear functionals
can be expressed as elements of ${\cal V}$ via the inner product.

It is easy to prove that in any inner product space the function $p_y$ defined by
\be 
p_y(x)=|(x,y)|\label{py}
\ee 
is a seminorm. The corresponding topology is the {\it weak topology} $\tau_0$. This topology has important properties.

{\bf Theorem}. The weak topology $\tau_0$ is a partial majorant in ${\cal V}$. If the inner product is non-degenerate the space is {\it separated (Hausdorff)}. Moreover $\tau_0$ is admissible.

When ${\cal V}$ is assigned the $\tau_0$ topology, it will be denoted by ${\cal V}_w$.

{\bf Remark}. The topology $\tau_0$ is not a majorant. Indeed a theorem, \cite{Bognar}, tells us that the topology $\tau_0$ is a majorant only if the space is finite-dimensional, which is not our case. In view of this, in the applications below we will have to deal only with partial majorants. Another question one could ask is whether $\tau_0$ is metrizable. Another theorem says, quite predictably, that if $\tau_0$ is metrizable it is also a normed partial majorant.

For later use we have to define the concept of bounded set. A subset $B$ is bounded if 
for any neighborhood $V$ of 0 there is a positive number $\lambda$ such that $B\subset \lambda V$. In terms of seminorms we can say that $B$ is bounded if all seminorms are bounded by some finite number in $B$. 

One may wonder why we do not use the seminorm $\|x\|= \sqrt{|(x,x)|}$ to define the topology. This can be done and the corresponding topology is called {\it intrinsic}, $\tau_{int}$. However such a topology does not guarantee continuity of all the functionals of the type $\varphi_y(x)=(x,y)$, see \cite{Bognar}. So one, in general, has to live with infinite many seminorms.

Another important question in dealing with topological vector spaces is the existence of a countable base of neighborhood. A base ${\cal B}$ of neighborhoods of the origin is a subset of all the neighborhoods of the origin such that any neighborhood in the given topology contains
an element of  ${\cal B}$. If the space is Hausdorff and the base is countable we say that the 
space satisfies the second axiom of countability, which is an important property because it permits us to use sequences (instead of filters) to study convergence.

Now, let us return to  ${\boldsymbol {\EF}}$ with the inner product $\langle\cdot |\cdot\rangle$ defined via (\ref{innerproduct}). Using it we can define an infinite set of seminorms as above and thereby the weak $\tau_0$ topology. In virtue of the preceding discussion 
${\boldsymbol {\EF}}$  becomes a topological vector space with a separated admissible topology.
We can also assume that the second axiom of countability holds for ${\boldsymbol {\EF}}$. This is
due to the fact that, apart from the $\e$ dependence, we can numerate the basis of all
possible states $\Omega_\alpha$. As for $\e$ we can discretize it, i.e replace it with
a sequence $\e_n$ tending to 0. In this way the index $\alpha$ is replaced by a discrete multi-index $\nu$ and we obtain a countable set of seminorms $p_{\nu}$. The neighborhoods of the origin defined by these seminorms form a countable basis. Finally, ${\boldsymbol {\EF}}$ with the $\tau_0$ topology is not a normed partial majorant, therefore it is not metrizable.

To stress that ${\boldsymbol {\EF}}$ is equipped with the $\tau_0$ topology we will use the symbol  ${\boldsymbol {\EF}}_w$.

We could stop at this point, remarking that, since the topology is admissible, any continuous functional can be expressed in terms ${\boldsymbol {\EF}}$. The $\tau_0$ topology is so `coarse' that it accommodates
simultaneously test states and distributions. However in ${\boldsymbol {\EF}}$ we can have a stronger topology. We say that a topology $\tau_1$ is stronger or finer than $\tau_2$ ($\tau_1\geq \tau_2$) if any open set in $\tau_2$ is an open set also in $\tau_1$. It is a theorem that if $\tau$ is
locally convex and stronger than $\tau_0$ it is also a partial majorant, which guarantees
continuity of the scalar product also wrt $\tau$. We will shortly introduce on ${\boldsymbol {\EF}}$ the {\it strong topology}. But to do so we need first to discuss the topology on the dual.

\subsection{The dual space}

Given a topological vector space ${\cal V}$ as above, the dual ${\cal V}'$ is the space of linear continuous functionals. Let us denote linear continuous functionals by $x',y',...$ and
their evaluation over a point $x\in {\cal V}$ by $x'(x), y'(x),...$. 

The {\it weak topology} over ${\cal V}'$ can be defined as follows: a sequence of linear continuous functionals $x'_n$ weakly converges to 0, if the numerical sequence $x'_n(x)$ converges to 0 for any $x\in {\cal V}$. 
Alternatively one can define a basis of neighborhoods of zero in ${\cal V}'$ as follows:  
\be 
U'_{\epsilon}(x_1,\ldots,x_r)=\big{\{}x'\in {\cal V}'\big{|}\, |x'(x_j)|\leq \epsilon ,\quad j=1,\ldots r\big{\}}\label{U'e}
\ee
for any subset $\{x_1,\ldots x_r\}$ in the family of finite subsets of ${\cal V}$. This topology turns ${\cal V}'$ into a locally convex topological vector space.

A subset $B'\in {\cal V}'$ is (weakly) bounded if for any neighborhood $U'_{\epsilon}$ as in (\ref{U'e})
there exist a positive number $\lambda$ such that $\lambda B'\subset U'_{\epsilon}$.

The space ${\cal V}'$ with the weak topology will be denoted ${\cal V}_w'$.
 
We can immediately transfer these concepts to the space ${\boldsymbol {\EF}'}$ of linear continuous functionals over ${\boldsymbol {\EF}}$, which is therefore itself a convex topological vector space. The space ${\boldsymbol {\EF}'}$ with  the weak topology will be denoted by ${\boldsymbol {\EF}'}_w$.

\subsection{The strong topology}

Using the weak topology on ${\cal V}'$ we can now define the {\it strong topology} on ${\cal V}$. The latter is defined as the uniform convergence topology on all weakly bounded subsets of ${\cal V}'$. This means that a sequence $x_n$ converges to $0$ in ${\cal V}$ if the numerical functions $x'(x_n)$ converge to zero uniformly for $x'$ in any bounded
subset $B$ of ${\cal V}'$. Alternatively we can define the strong topology by means of a basis of neighborhoods of 0. A neighborhood $V_{\e}$ of 0 is defined by
\be 
V_{\epsilon}= \bigl{\{}x \in {\cal V}\big{|} {\rm sup}_{x'\in B} |x'(x)|<\epsilon\bigr{\}}\label{Ve}
\ee
for any $\epsilon$ and any bounded set $B\subset {\cal V}'$. ${\cal V}$ equipped with the strong topology will be denoted by ${\cal V}_s$.

We recall that when ${\cal V}$ is assigned the weak $\tau_0$ topology, for any continuous functional  $x'\in{\cal V}'$ we have $x'(x)= (x,y)$ for some $y\in {\cal V}$. This is generically not true for the dual of ${\cal V}$ when ${\cal V}$ is equipped with the strong topology. The dual of  ${\cal V}_s$ is generally larger than ${\cal V}'$. In fact a theorem says that any seminorm which is lower semicontinuous in ${\cal V}_w$ is continuous in 
${\cal V}_s$, in other words there are more continuous seminorms in ${\cal V}_s$ than in ${\cal V}_w$. Qualitatively speaking, this means that there is in ${\cal V}_s$ a smaller number of convergent sequences than in ${\cal V}_w$, which implies that there are more continuous functionals.

The dual of ${\cal V}_s$ will be denoted by ${\cal V}'_s$. For completeness we add that it can itself be equipped with a {\it strong topology} as follows: a neighborhood $V'_{\epsilon}$ of 0 in ${\cal V}'_s$ is defined by
\be 
V'_{\epsilon}= \big{\{} x' \in {\cal V}'\big{|} {\rm sup}_{x\in B}|x'(x)|<\epsilon\big{\}}\label{V'e}
\ee
for any $\epsilon$ and any bounded set $B\subset {\cal V}_s$. ${\cal V}'_s$ equipped with the strong topology will be denoted also as ${\cal V}'_{ss}$.

We can immediately transfer these concepts to the space ${\boldsymbol {\EF}}$ and its duals. The space ${\boldsymbol {\EF}'}$ with the weak topology will be denoted by  ${\boldsymbol {\EF}'}_w$ and ${\boldsymbol {\EF}}$ with the strong topology will be denoted by ${\boldsymbol {\EF}}_s$. The dual of the latter will be denoted with the symbol ${\boldsymbol {\EF}}'_s$.

\subsection{`Richness' of the space of test states}

The space ${\boldsymbol {\EF}}$ equipped with the weak or strong topology will be our {\it space of test states}. The dual of the latter, i.e. ${\boldsymbol {\EF}}'$ or ${\boldsymbol {\EF}}'_s$ will be our space of {\it generalized states} or {\it distributions}. We can equip the latter with the weak or strong topology according to the needs.
 
As in ordinary distribution theory we have to verify that ${\boldsymbol {\EF}}$ is a rich enough filter so that no regular behaviour can escape through it. We first remark that the cardinality of the basis ${\boldsymbol{\Omega}}_\alpha$ with fixed $\e$ is the same as the cardinality of the Fock space states $\boldsymbol {\cal F}$. If we include the $\e$ dependence the cardinality of ${\boldsymbol {\EF}}$ is larger. Let us also add that in the representation (\ref{qualitative}) of correlators, any kind of inverse integer powers of $s$ appear in the IR, and any kind of half integer power of $s$ appear in the UV. This is what our intuition would suggest to guarantee completeness.
 
More formally, let us compare the situation here with sec.2.4. To be able to claim that ${\boldsymbol {\EF}}$ is {\it rich  enough} we must show that a state that annihilates the full  ${\boldsymbol {\EF}}$  can only be 0. To see this let us consider a generic finite linear combination of states ${\boldsymbol{\Omega}}_\alpha$, say $\boldsymbol{\Upsilon}$,
and suppose that 
\be 
\langle  \boldsymbol{\Upsilon}| {\boldsymbol{\Omega}}_\alpha\rangle=0, \quad\quad \forall\,
{\boldsymbol{\Omega}}_\alpha\in{\boldsymbol {\EF}}\label{richness}
\ee
If such a state $\boldsymbol{\Upsilon}$ were to exist it would mean that the inner product (\ref{innerproduct}) is degenerate. As far as we can exclude the degeneracy of the inner product we conclude that ${\boldsymbol {\EF}}$ is a rich enough space of test states.

\section{Some conclusions and comments}

In the light of the construction presented in the previous section, we would like now to return to the questions raised in the introduction and in section 4. The state $\Lambda_\e$ can be accommodated in the dual of ${\boldsymbol {\EF}}$. This follows from (\ref{innerproduct}). Let us keep $\Omega_\beta$ fixed while $\Omega_\alpha$ spans ${\boldsymbol {\EF}}$. A discontinuity
of $\langle \Omega_\alpha\, Bc\partial c\,\Lambda_\e\rangle$ would imply a discontinuity of the inner product
in the $\Omega_\alpha$ entry on the LHS. But this contradicts the fact that in the $\tau_0$ topology the inner product is separately continuous in the two entries. Therefore
$\langle \Omega_\alpha\, Bc\partial c\,\Lambda_\e\rangle$  is continuous, i.e. it belongs to ${\boldsymbol {\EF}}'$ (for any value of $\e$ including 0). As a consequence of the construction in sec. 6, it also belongs to ${\boldsymbol {\EF}_s}'$.

This is probably the simplest way to think of $\Lambda_0= \lim_{\e\to 0} \Lambda_\e$ as a distribution. In analogy with the example ${\rm x}^{-1}$ in sec.2.3 we call this the {\it principal value} regularization of
$\Lambda_0$, i.e of $\frac 1{K+\phi_u}$. For the same reason we can also conclude that 
${\cal A}_0$ is the null distribution in ${\boldsymbol {\EF}}'$ (see also the discussion below on this point). These conclusions hinge upon the structure of ${\boldsymbol {\EF}}$, and in particular on the fact that all the test states correlators used to define the inner product are represented, via (\ref{qualitative}), by integrands $F(s)$ that decrease at least as fast as $\frac 1{s^2}$ in the IR. 

With the above principal value regularization it is not possible to capture the contribution (if any) from the `pointlike' support of ${\cal A}_\e$ for $\e\to 0$, mentioned at the beginning of sec. 5. This question is important even regardless the invertibility of $K+\phi_u$, for, as we have mentioned in the introduction, it is believed that the limit: $\lim_{t\to\infty} e^{-t(K+\phi_u)}$, represents a sliver-like projector. It would be important to find an adequate mathematical representation of such an object (if it exists). We would now like to explore the possibility to capture 
such a delta-like object in the functional analytic framework introduced above. 

The term ${\cal A}_0$ is of the type (\ref{deltalim}) or (\ref{deltalim3}). We recall
that the latter is actually 0, see (\ref{deltalim4}). Let us write
\be
{\cal A}_0 = \lim_{\e\to 0} \,\e \int_0^\infty dt\, e^{-t(K+\phi_u+\e)}\label{Ae1}
\ee
Since, what is relevant here is the eigenvalue of ${\cal K}_u$ near 0, we can think of replacing $K+\phi_u$ with its eigenvalue and integrating over it to simulate the path integration. The eigenvalue of ${\cal K}_u$ is a function of some spectral parameter $\kappa$. So we replace $K+\phi_u$ by $\kappa^a$, with $a>0$ (it can only be a power of $\kappa$ since it must vanish for $\kappa\to 0$). Then we have
\be
{\cal A}_{\e}=\e\int_0^\infty dt e^{-t\e} \int_0^m d\kappa \, e^{-t \kappa^a}\approx 
\e\int_0^\infty dt\, e^{-t\e}\,t^{-\frac 1a} 
\sim  \left\{\begin{matrix}
 \e^{\frac 1a} \log\e &\quad &a\leq 1\\
\e^{\frac 1a} & \quad &a>1\end{matrix}\right.\label{simul}
\ee
where $m$ is an arbitrary small finite number that does not affect the result.

Thus ${\cal A}_{0}=0$, at least according to this heuristic treatment. This approach understands a sort of strong operator topology. In order to capture a nonzero contribution in ${\cal A}_\e$
one must allow for string states with corresponding integrands in (\ref{qualitative}) that tend to a constant value in the IR, when the factor $e^{-\e t}$ is suppressed. This means that the
$\e\to 0$ limit for these states is discontinuous. Therefore they can hardly be considered test states. In conclusion, the empirical formula (\ref{Ainverseobstr}) does not seem to be fit to capture the delta-function-like content (if any) of $\frac 1{K+\phi_u}$. In addition, the examples in sec. 2.3 suggest that we would have to evaluate
$(K+\phi_u)^\lambda$ with complex $\lambda$ and then proceed like for ordinary distributions. Unfortunately we are unable to evaluate such an expression using the Schwinger representation.

In general one expects that there are several different ways to represent a regularized inverse of $K+\phi_u$, in analogy with the inverse of $x$ in section 2. But the formalism we can avail ourselves of has at present technical limitations. The only sensible course (at least for the time being) is to use the principal value regulated inverse defined at the beginning of this section. This is what we understand from now on.
\vskip 1cm
We can now return to the three conditions (a,b,c) of section 4. When proving the equation of motion one has to use the regularized inverse of $K+\phi_u$. We have already remarked that in such a way there is no violation to the equation of motion. Simultaneously with condition (c) also condition (b) is satisfied, because the existence of $\frac 1{K+\phi_u}$ means precisely that we can compute it against any test function. As for condition (a), it is of a different nature, it arises from a different requirement: if the homotopy operator $\frac B{K+\phi_u}$ applied to a normalized (perturbative) state were to yield a normalized state, the cohomology of ${\cal Q}_{\psi_u}= Q+\{\psi_u,~\cdot~\}$ would be trivial, and the solution $\psi_u$ would not represent a lump. As shown in \cite{BMT} there are more than one indication that this is not the case: $\tr \frac 1{K+\phi_u}$ is  infinite and we have just shown that $\frac 1{K+\phi_u}$ must be understood as a distribution. This is enough to reassure us that
$\frac B{K+\phi_u}$ is not a good homotopy operator. 

\subsection{Final comments}

In this paper we have proposed a framework in which objects such as the inverse of $K+\phi$ can be consistently defined. We have done it by introducing  a locally convex topological vector space of string states, with either weak and strong topology, and using the dual space as a distribution space. The inverse of $K+\phi$ turns out to be an object in this space of functionals and to correspond to a regularization we have referred to as `principal value' regularization. Although we have not done it in detail, also the inverse of $K$ can be treated in a similar way (i.e. using matter as a regulator). Admittedly our approach has been very concrete and case-oriented. For instance, basing the topology on the the inner product (\ref{innerproduct}) seems to strongly limit the power of the formalism. A more general approach should be possible along the lines of \cite{Schwartz} (which however deals only with finite dimensional vector spaces). It is clear that the basic space is $\boldsymbol {\cal F}$ introduced in section 3 (or rather its generalization including ghosts and zero modes). Maps from the string world-sheet to this space represent string configurations. Therefore the latter space of maps and its topologies is the real thing to be studied. In this framework the Fock space states correspond to constant maps, and it is understandable that they may be of little use as test states. The hard problem is the definition of the topology in the above space of maps and the duality rule. In our construction in section 5 and 6 both problems were solved thanks to the knowledge of the exact relevant partition function of \cite{Witten2}. In general one has to make do without it. This seems to be the true challenge.

\vskip 1cm
{\bf Acknowledgments.} 
 
L.B would like to thank Gianni Dal Maso for sharing with him his expertise on functional analysis. We would like to thank D.D.Tolla for discussions and for our using material from previous joint papers.
The work of L.B. and S.G. was supported in part by the MIUR-PRIN contract 2009-KHZKRX.

\section*{Appendix}
\appendix

\bigskip
In this Appendix we would like to briefly discuss the spectrum of operator $K_1^L$. The latter operator is defined by 
\be 
K_1^L=\frac 12 K_1 - \frac 1{\pi} \left({\cal L}_0+{\cal L}_0^\dagger\right)\label{K1L}
\ee
where $K_1=L_1+L_{-1}$ and
\be 
{\cal L}_0+{\cal L}_0^\dagger= 2 L_0 +\sum_{n=1}^\infty \ell_{2n}\left(L_{2n}+L_{2n}^\dagger\right),
\quad\quad \ell_{2n}= \frac{2(-1)^{n+1}}{4n^2-1}\label{L0L0dag}
\ee
$L_n$ represent the total (matter+ghost) Virasoro generators. For the sake of simplicity we restrict ourselves here to the matter part since it is enough to appreciate the complexity of the problem. In terms of oscillators $a_n,a_n^\dagger$, $n=1,2,...$ (we forget $a_0$ because we are considering 0 momentum states), we can write in compact notation
\be
K_1= a^\dagger\cdot F \, a, \quad\quad {\cal L}_0+{\cal L}_0^\dagger= a^\dagger\, A  a^\dagger + a^\dagger C\, a+a\, A\, a\label{A1}
\ee
where $F, A, C$ are $\infty \times\infty$ symmetric numerical matrices. Their explicit expressions can be found in (\cite{RSZ},\cite{BMST1}). We recall here only the properties
\be
AF+FA=0,\quad\quad [C,F]=0\label{A2}
\ee
which imply that 
\be
[K_1, {\cal L}_0+{\cal L}_0^\dagger]=0\label{A3}
\ee
The matrix $C$ and the twisted matrix $\tilde A$ can be diagonalized on the basis $v_n(\kappa)$ of $F$ eigenvectors
\be 
\sum_{m=1}^\infty F_{nm} v_m(\kappa)= \kappa \, v_n(\kappa)\label{A4}
\ee
The relevant eigenvalues ${\mathfrak c}(\kappa)$ and ${\mathfrak a}(\kappa)$ can be found again in (\cite{RSZ},\cite{BMST1}). We have in particular
\be 
[K_1, a^\dagger\! \cdot v(\kappa)] = \kappa \, a^\dagger\! \cdot v(\kappa)\label{A5}
\ee
This means that $a^\dagger \cdot v(\kappa)|0\rangle$ is an eigenstate of $K_1$ with eigenvalue $\kappa$. Due to (\ref{A2},\ref{A3}) any state of the form
\be 
f({\cal L}_0+{\cal L}_0^\dagger) a^\dagger \cdot v(\kappa)|0\rangle, \quad\quad {\rm or}
\quad\quad g(a^\dagger A a^\dagger)  a^\dagger \cdot v(\kappa)|0\rangle\label{A6}
\ee
where $f,g$ are arbitrary analytic functions, are also eigenstates of $K_1$ with eigenvalue $\kappa$. Differently from what happens for the matrices $F,\tilde A$ and $C$, which have common eigenvectors, the situation for the Fock space operators $K_1$ and ${\cal L}_0+{\cal L}_0^\dagger$ is much more complicated: there is an infinite degeneracy corresponding to each eigenvalue $\kappa$ of $K_1$. What one should do next is extract from the infinite families
(\ref{A6}) the eigenvectors of ${\cal L}_0+{\cal L}_0^\dagger$ and calculate the corresponding eigenvalues. Only in this way will one be able to compute the eigenvalues of $K_1^L$. Needless to say one should also consider the ghost part of $K_1^L$ (for which the results of {\cite{BMST1,BMST2} may be instrumental). Unfortunately these problems are still waiting for a solution. 


\end{document}